\documentclass[a4paper]{jpconf} 

\usepackage{graphicx}
\usepackage{amsmath}
\usepackage{amsfonts}
\usepackage{amssymb}

\newcommand{\zeroth}{0$^\mathrm{th}\,$}
\newcommand{\first}{1$^\mathrm{st}\,$}

\begin{document}
\title{Parallax in angular sensitive powder diffraction tomography}

\author{Peter Modregger$^\mathrm{1,2}$,  Ahmar Khaliq$^\mathrm{1,2}$ and Felix Wittwer$^\mathrm{1,2}$}
\address{$^\mathrm{1}$Physics department, University of Siegen, Siegen, Germany}
\address{$^\mathrm{2}$Centre for X-ray and Nano Science, CXNS, DESY, Hamburg, Germany}
\ead{peter.modregger@uni-siegen.de}

\begin{abstract}
While a few methods for the determination of depth-resolved strain distributions each with inherent limitations are available, tomographic reconstruction has been applied to this problem in only a limited sense. One of the challenges was the potential impact of geometric parallax, which constitutes a non-negligible lateral offset of diffraction information arising from different sample depths at the detector. Here, the effect of parallax was investigated and two main results have emerged. First, the impact of parallax was found to be additive to other offset contributions, which implies a straightforward correction. Second, for tomographic scans utilizing a full 360\textdegree{} rotation parallax has been found to have no impact on reconstructions of angular information.
\end{abstract}

\section*{Introduction}

The determination of depth-resolved strain and stress distributions in bulk engineering materials remains a challenging task. In general, there are two types of methods available: X-ray diffraction with conical slits~\cite{Staron2014}, requiring challenging to manufacture optical elements, and energy dispersive X-ray diffraction~\cite{Apel2018}, which utilizes energy dispersive X-ray detectors. Further, both of these methods share the disadvantage of providing highly anisotropic gauge volumes of 1:10 and higher.

The possibility for tomographic reconstruction of the local distribution of the six strain tensor components has been discussed for nearly two decades~\cite{Korsunsky2006}. In 2015, Lionheart and Withers have explored this challenge from a mathematical point of view~\cite{Lionheart2015}. They have found that the diffraction information from scans of six carefully chosen rotation axes provided {\em sufficient} but not necessary data to reconstruct the strain tensor components. Indeed, they purposefully have left open the potentiality for using a smaller number of axes. Korsunksy et al. have demonstrated the tomographic reconstruction of rotationally symmetric strain~\cite{Korsunsky2006} as well as the averaged strain (i.e., one component)~\cite{Korsunksy2011}.

In the following, a particular challenge of strain tomography has been investigated: the geometric parallax arising from a non-negligible sample thickness. First, the theoretical framework is presented, then details of a pilot experiment are given. Afterwards, numerical simulations have been used to quantify the impact of parallax on tomographic reconstructions of angular sensitive diffraction data.

\section*{Parallax}

\begin{figure}[htbp]
\begin{center}
\includegraphics[width=0.5\textwidth]{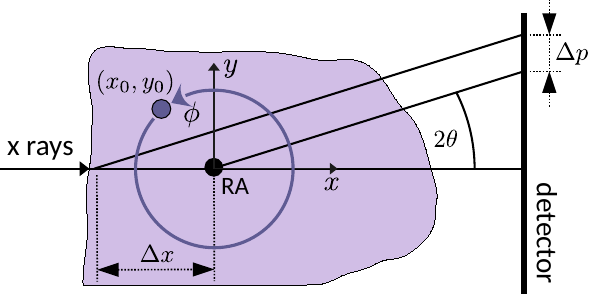}
\end{center}
\caption{Definition of parallax. The location of the rotation axis (RA) is chosen the origin of the laboratory reference frame.}
\label{fig:parallax}
\end{figure}

In the present context, parallax refers to a geometrical, lateral offset of beams at the detector that were diffracted at different sample depths (see Fig.~\ref{fig:parallax}). The discussion has been limited to a plane orthogonal to the rotation axis (i.e., a 2D slice). Extension to three dimensions is straight forward. Further, the investigation has been limited to one specific Bragg angle $\theta$ (i.e., a single diffraction ring), which was without loss of generality as long as the diffraction curves from different Bragg angles do not overlap. In addition, isotropic diffraction intensities during tomographic rotation have been assumed.

During rotation x rays diffracted at the position of the rotation axis will always hit the detector at the same spot. Thus, the rotation axis served as a convenient choice for the origin of the laboratory reference frame. Since the sample is scanned along the $y$ direction the  parallax offset $\Delta p$ of a specific sample feature only depends on the horizontal offset $\Delta x$ and is given by (see Fig.~\ref{fig:parallax})
\begin{equation}\label{eq:parallax}
\Delta p = \Delta x \tan2\theta.
\end{equation}
For the setup described above, a representative scattering angle of $2\theta = 14$\textdegree{} and a horizontal offset of $\Delta x=0.5$~mm implies an offset of $0.16$~mrad or 0.8~pixels. Compared to a desired accuracy of diffraction curve offsets in the micro-radian regime, parallax may constitute a significant influence warranting further investigation.

For this, a model for contrast formation similar to the one presented in~\cite{Lionheart2015} was used. Each voxel in the final tomographic reconstruction was associated with a local diffraction curve $f(x,y,\phi,\Delta\theta)$ that depends on the spatial positions $(x,y)$, the rotation angle $\phi$ and on the Bragg angle offset $\Delta\theta$. The observable diffraction pattern at the detector, $F(t,\phi,\Delta\theta)$ with $t$ the lateral offset from the rotation axis, was then given as a projection of $f$, $\mathcal{R}[f]$, onto the detector,~\cite{Kak1988}
	\begin{equation}\label{eq:F}
	F(t,\phi,\Delta\theta) = \mathcal{R}[f] = \int\!\!dxdy\, \delta_D(t-
	x\cos\phi-y\sin\phi)\,f(x,y,\phi,\Delta\theta).
	\end{equation}
Here, the projection was conveniently written via Dirac's $\delta_D$-distribution. Thus, neglecting effects of absorption within the sample, the observable diffraction patterns were simply given as the sum of local diffraction curves along the beam. 

Diffraction curves were analyzed in terms of the \zeroth and \first moment~\cite{James2006} according to the suggestion in~\cite{Lionheart2015}. The following demonstrated that the un-normalized \first moments form a line integral and that additive, local contributions to the \first moment add up. The \zeroth moment 
	\begin{equation}\label{eq:zeroth_moment}
	M_0[f] = \int\!\!d\Delta\theta\, f(x,y,\phi,\Delta\theta)
	\end{equation}
corresponded to the total diffracted intensity. The un-normalized \first moment was given as 
	\begin{equation}\label{eq:un_first_moment}
	M_1[f] = \int\!\!d\Delta\theta\, \Delta\theta\, f(x,y,\phi,\Delta\theta),
	\end{equation}
while the normalized \first moment, $\bar M_1$, was calculated by
	\begin{equation}\label{eq:norm_first_moment}
	\bar M_1[f] = \frac{M_1[f]}{M_0[f]}.
	\end{equation}
For the local diffraction curves $f(x,y,\phi,\Delta\theta)$, the latter corresponded to an angular offset of diffraction curves, which was given as the sum of a parallax contribution $\Delta\theta_p$ and other contributions $\Delta\theta_s$ such as strain:
	\begin{equation}\label{eq:local_offsets}
	\bar M_1[f(x,y,\phi,\Delta\theta)] = \Delta\theta_p + \Delta\theta_s.
	\end{equation}
Here, it was assumed that the \first moment of diffraction curves without external contributions is zero.

Combining the definitions of the projection of local diffraction curves,  eq.~(\ref{eq:F}), and the un-normalized \first moment ,eq.~(\ref{eq:un_first_moment}), yielded the un-normalized \first moment of the detectable diffraction curves:
	\begin{equation}
	M_1[F] = \int\!\!d\Delta\theta\, \Delta\theta \int\!\!dxdy\, \delta_D(t-	
	x\cos\phi-y\sin\phi)\,f(x,y,\phi,\Delta\theta).
	\end{equation}
Using the definition of the normalized \first moment, eq.~(\ref{eq:norm_first_moment}) and the local angular offsets, eq.~(\ref{eq:local_offsets}) provided 
	\begin{equation}
	M_1[F] = \int\!\!dxdy\, \delta_D(t-
	x\cos\phi-y\sin\phi)\,M_0[f] (\Delta\theta_p + \Delta\theta_s),
	\end{equation}
which demonstrated that the un-normalized \first moments of local diffraction curves form a simple line integral. Please note that this does not apply to the corresponding normalized \first moments. Further, this showed that the projections of parallax and other contributions to local offsets add up, i.e.
	\begin{equation}\label{eq:aaa}
	M_0[F]\,\bar M_1[F] = \mathcal{R}[M_0[f]\Delta\theta_p] + \mathcal{R}
	[M_0[f] \Delta\theta_s].
	\end{equation}
The latter implied that the effect of parallax on tomographic reconstructions can be investigated independently from other contributions.

During rotation the parallax experienced by a specific sample detail located at the position $(x_0,y_0)$ (see also Fig.~\ref{fig:parallax}) was given as 
	\begin{equation}\label{eq:point_parallax}
	\Delta\theta_p = (x_0\cos\phi + y_0\sin\phi) \frac{\tan 2\theta}{z}
	\end{equation}
with $z$ the distance between the sample and the detector. In the case of utilizing a full 360\textdegree{} rotation, the averaged parallax experienced by any sample point was
	\begin{equation}\label{eq:average}
	\langle\Delta\theta_p\rangle = (x_0\langle\cos\phi\rangle +
	y_0\langle\sin\phi\rangle) \frac{\tan 2\theta}{z} = 0,
	\end{equation}
where the symbol $\langle\cdot\rangle$ indicated the average. Thus, using a full 360\textdegree{} rotation implied that parallax would have no effect on tomographic reconstructions.

\section*{Experiment}

\begin{figure}[htbp]
\begin{center}
\begin{tabular}{cc}
\includegraphics[width=0.51\textwidth]{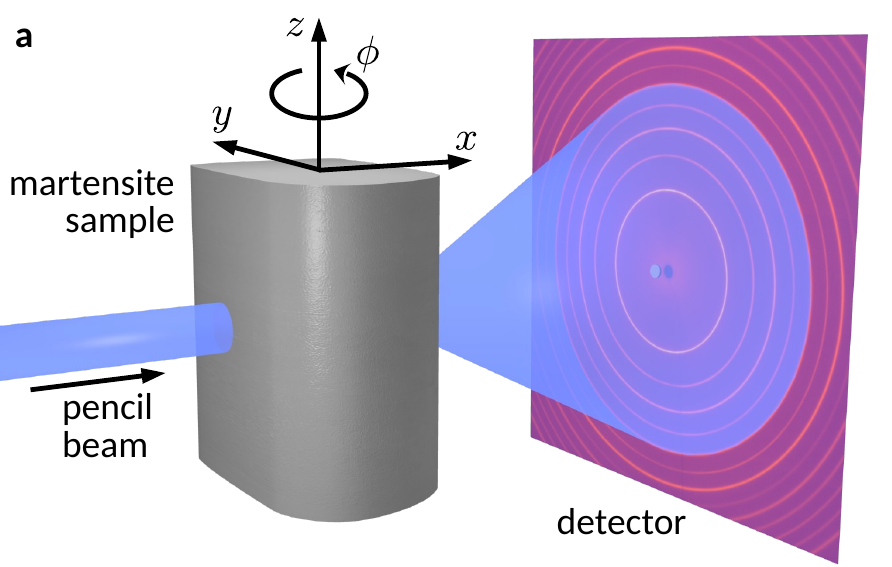} &
\includegraphics[width=0.42\textwidth]{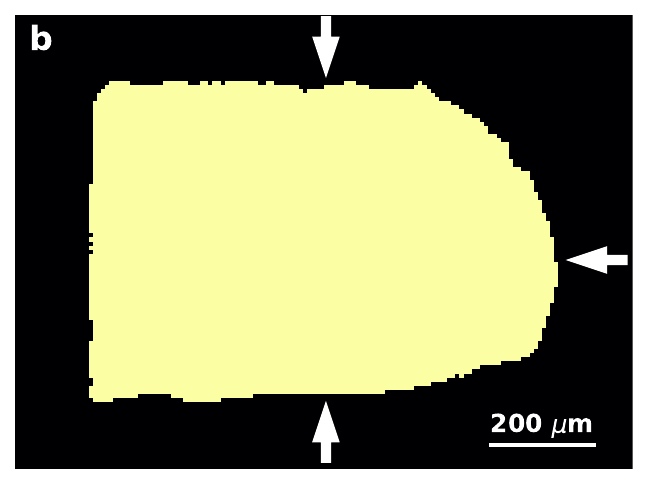} 
\end{tabular}
\end{center}
\caption{(a) Sketch of the experimental setup. (b) Sample shape retrieved from tomographic reconstruction of total diffraction intensity. The arrows indicate the surfaces, which were shot peened.}
\label{fig:setup}
\end{figure}

This pilot experiment was carried out at the P21.2 beamline~\cite{Hegedus2019} of the synchrotron radiation facility PETRA III in Hamburg (Germany). A sketch of the setup is shown in Fig.~\ref{fig:setup}a. A photon energy of 68~keV was selected and the lateral beam dimension were 8~$\mu$m $\times$ 3~$\mu$m. The sample was 1~mm martensitic steel, which was shot-peened from three sides as indicated in Fig.~\ref{fig:setup}b. Diffraction rings were collected by a VAREX XRD4343CT detector with a pixel size of 150~$\mu$m approximately 0.8~m downstream of the sample. The discussion in this paper was limited to the martensite (321) reflection, which corresponded to a Bragg angle of $\theta = 6.839$\textdegree{}. The scan consisted of 200 horizontal translation and 200 rotation steps, the latter over 360\textdegree{}, resulting in the acquisition of 40,000 diffraction patterns, which took approximately 5~h. Diffraction from a Ce0$_2$ calibrant powder sample and geometry fitting by pyFAI~\cite{pyfai} were used to calibrate the detector angles.

\section*{Results}

\begin{figure}[htbp]
\begin{center}
\includegraphics[width=0.33\textwidth]{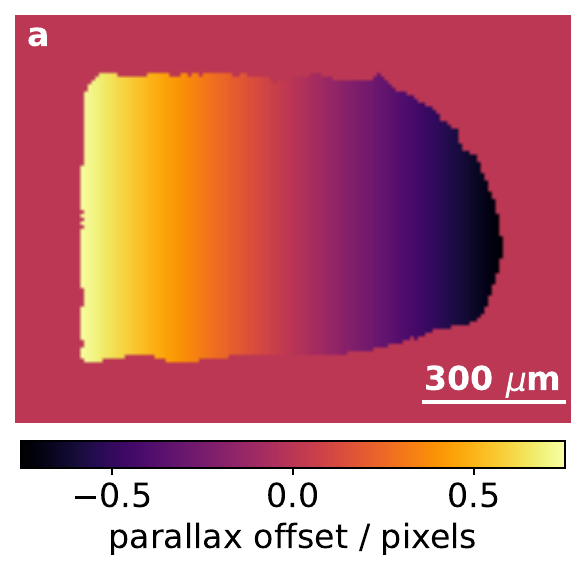} 
\includegraphics[width=0.33\textwidth]{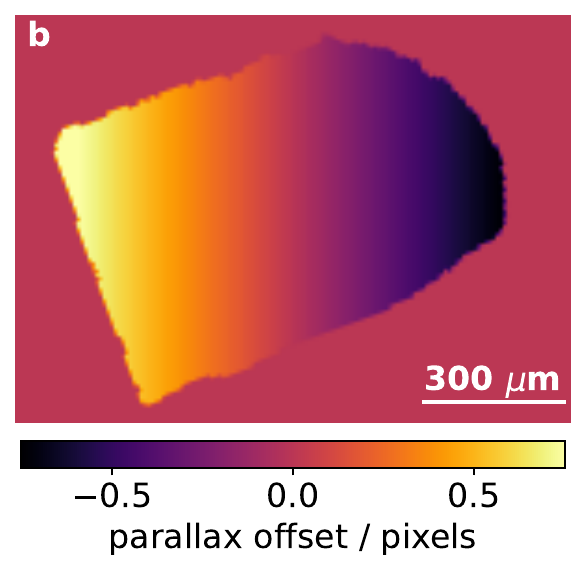} \\
\includegraphics[width=0.32\textwidth]{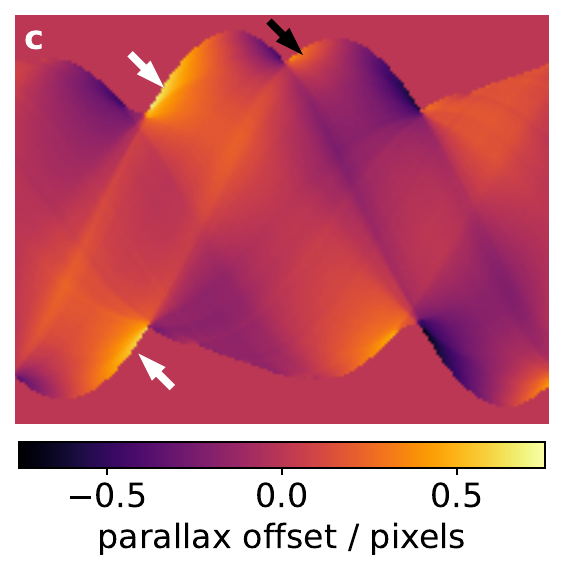} 
\includegraphics[width=0.33\textwidth]{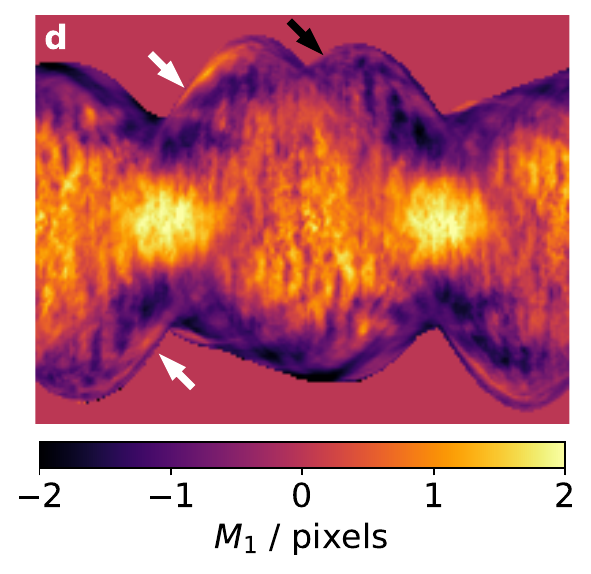}
\end{center}
\caption{Parallax example. (a) Theoretical parallax offset at a rotation angle of $\phi=0$\textdegree{} for each sample position expressed in pixels. (b) Same as (a) but at a rotation angle of $\phi=20$\textdegree{}. (c) Sinogram corresponding to (a). (d) Experimental sinogram of a shot-peened martensite sample. The white (black) arrows indicate some points of (dis-)agreement between theory (b) and experiment (c).}
\label{fig:parallax_example}
\end{figure}

Numerical simulations have been used to investigate the impact of parallax on potential tomographic reconstructions of the \first moments of observable diffraction curves. Fig.~\ref{fig:parallax_example}a shows an example for the spatial distribution of the parallax offset within the sample at the projection angle $\phi=0$. As the parallax explicitly depends on the rotation angle $\phi$ in eq.~(\ref{eq:point_parallax}), the projection of the parallax 
	\begin{equation}
	\bar M_1[F] = \frac{\mathcal{R}[M_0[f]\Delta\theta_p]}{\mathcal{R}
	[M_0[f]]},
	\end{equation}
which was obtained from eq.~(\ref{eq:aaa}) with $\Delta\theta_s=0$, did {\em not} constitue a straight forward Radon transform. For a quasi homogeneously and isotropically diffracting sample (i.e., $M_0[f(x,y,\phi] = \mathrm{const}$), the equation simplified to 
	\begin{equation}\label{eq:parallax_sino}
	\bar M_1[F] = \frac{\mathcal{R}[\Delta\theta_p]}{\mathcal{R}
	[s]},
	\end{equation}
with $s$ the path length within the sample boundaries. A similar equation was also reported in~\cite{Korsunsky2006}. The sinogram of the parallax contribution eq.~(\ref{eq:parallax_sino}) was calculated by rotating the sample shape, adding the parallax distribution and subsequent summing. The resulting sinogram is shown in Fig.~\ref{fig:parallax_example}b.

This numerical expectation was compared to the experimental result, the latter shown in Fig.~\ref{fig:parallax_example}c. The sample for the experiment was shot-peened from three sides (corresponding to top, right and bottom in Fig.~\ref{fig:parallax_example}a), which introduced strong residual strains in the sample and overshadowed the contribution of parallax by around one order of magnitude. Nevertheless, the correlation coefficient between simulation and experiment was still $r=0.3$, which was taken as modest evidence for the viability of these investigations. In the future, the experiment should be repeated with an unstrained sample for a more rigorous validation.

\begin{figure}[htbp]
\begin{center}
\includegraphics[width=0.45\textwidth]{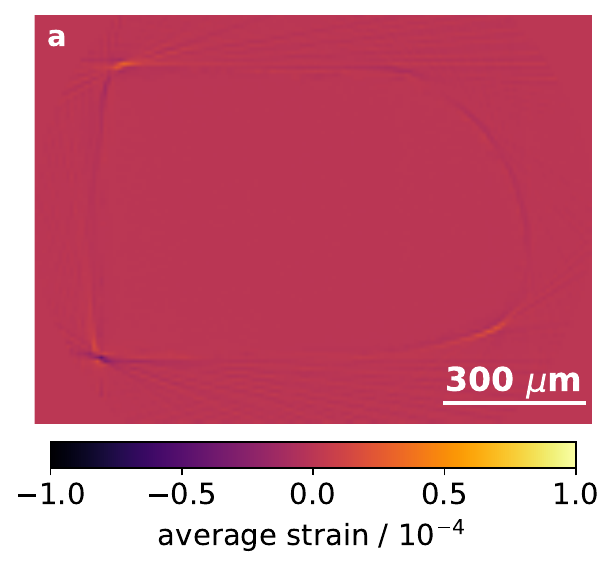}
\includegraphics[width=0.45\textwidth]{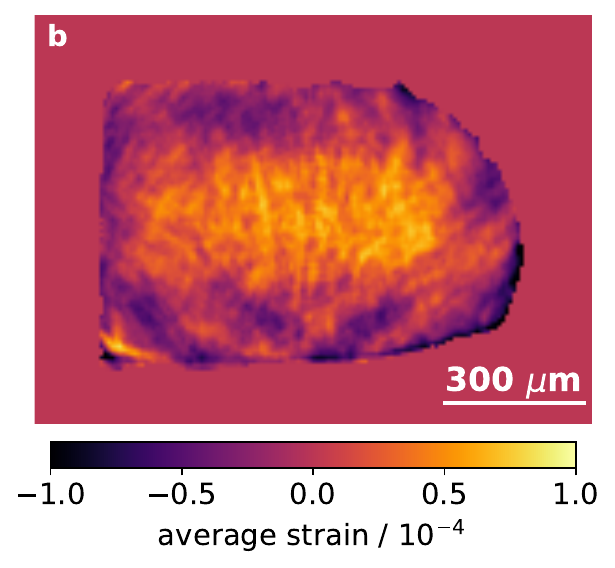}
\end{center}
\caption{Tomographic reconstructions. (a) Reconstruction of the parallax offset incorrectly treated as strain (i.e., tomographic reconstruction of Fig.~\ref{fig:parallax_example}b). (b) Tomographic reconstruction of the average strain of the martensite sample according to~\cite{Korsunksy2011}.}
\label{fig:parallax_rec}
\end{figure}

Finally, the prediction of zero influence of parallax to tomographic reconstructions utilizing full 360\textdegree{} rotation according to eq.~(\ref{eq:average}) was investigated numerically. To this end, the suggestion of reconstructing the average strain per voxel as presented in~\cite{Korsunksy2011} by a simple inverse Radon transform has been used. Fig.~\ref{fig:parallax_rec}a shows the reconstruction of the parallax sinogram (Fig.~\ref{fig:parallax_example}b), which was incorrectly treated as originating from strain. Apart from the edges of the sample, which was due to partial volume effects, the reconstruction was zero as expected. The tomographic reconstruction of the experimental sinogram (i.e., Fig.~\ref{fig:parallax_example}c) also showed an expected outcome: compressive average strain at the edges of the sample and tensile strain within the bulk. It was clear that the influence of parallax on the reconstruction was orders of magnitude smaller than effects from strain.

\section*{Conclusions}

The influence of parallax on tomographic reconstruction of angular data from powder diffraction has been investigated. Two key insights were reported. First, the influence of parallax is additive to other contributions to angular offsets of diffraction curves. This implied that a parallax correction of experimental sinograms is straightforwards Second, parallax has a negligible impact on tomographic reconstructions for scans utilizing a full 360\textdegree{} rotation.

\section*{Acknowledgements}

We acknowledge DESY (Hamburg, Germany), a member of the Helmholtz Association HGF, for the provision of experimental facilities. Parts of this research were carried out at PETRA III and we would like to thank Ulrich Lienert for assistance in using beamline P21.2.

\section*{References}

\bibliographystyle{unsrt}

\end{document}